# Multi-octave spectrally tunable strong-field Terahertz laser

Carlo Vicario[1], Andrey V. Ovchinnikov[2], Oleg V. Chefonov[2] and Christoph P. Hauri[1,3,|*]

[1]Paul Scherrer Institute, SwissFEL, 5232 Villigen PSI, Switzerland.

[2] Joint Institute for High Temperatures of RAS, Izhorskaya st. 13 Bd. 2, Moscow 125412, Russia.

[3]École Polytechnique Fédérale de Lausanne, 1015 Lausanne, Switzerland.

*Correspondence to: christoph.hauri@psi.ch

**The ideal laser source for the emerging research field of nonlinear Terahertz (THz) spectroscopy should offer radiation with a large versatility and deliver both ultra-intense multi-octave spanning single-cycle pulses and user-selectable multi-cycle pulses at narrow linewidth. The absence of such a table-top source has hampered advances in numerous THz disciplines including imaging, nonlinear photonics and spectroscopy, selective out-of-equilibrium excitation of condensed matter and quantum systems. Here we introduce a highly versatile table-top THz laser platform providing single-cycle GV/m transients as well as spectrally narrow pulses tunable in bandwidth and central frequency across 5 octaves with hundreds of MV/m field strength. The compact scheme is based on optical rectification of a temporally modulated laser beam in organic crystals. It allows for the selection of THz oscillation cycles from 1 to >50 and central frequency tuning range from 0.5 to 7 THz by directly changing the modulation period of the driving laser. The versatility of the THz source is demonstrated by providing a broadband 5-octave spanning spectrum as well as a spectrally narrow line tunable across the full optical rectification phase-matching band with a minimum width of $\Delta\nu\approx30$ GHz, corresponding to $\Delta E=0.13$ meV and $\lambda^{-1}=1.1$ cm$^{-1}$. The presented table-top source shows performances similar or even beyond to that of a large-scale THz electron accelerator facility but offering in addition versatile multi-color and advanced femtosecond pump-probe opportunities at ultralow timing jitter.**

**Introduction**

Terahertz radiation represents an ideal stimulus for controlling matter in a more advanced way compared to an optical light stimulus. While optical light predominantly interacts with valence electrons, Terahertz (THz) radiation, defined as the frequency region between 0.1-10 THz, allows direct and selective access to the numerous low-energy excitations including lattice vibrations, spin waves, molecular rotations, quantum states and the internal excitations of bound electron-hole pairs [1-7]. The recent advances in the available THz brightness also permit non-resonant control over matter where the THz stimulus acts as impulsive excitation [8-10]. While the selective control requires narrowband THz radiation the impulsive excitation relies on intense single-cycle THz pulses carrying a (multi-)octave spanning spectrum.

The ideal source for linear and nonlinear THz spectroscopy should provide the experimenter the free control on the number and the period of the field oscillations, from 1 to many cycles, and correspondingly on the spectral bandwidth from narrowband to broadband, and to apply highest possible field strength up to GV/m. While intense single-cycle pulses with field strength of GV/m have become recently available [11], the production of narrow bandwidth pulses at high field strength (>10 MV/m) which are tunable across several octaves has remained a formidable challenge. Intense THz radiation confined in a narrow spectral line is a necessity for ground-breaking investigations in imaging, light-induced electron acceleration and nonlinear excitation of resonant modes.

In this letter we demonstrate an advanced scheme which provides both intense laser-based THz radiation adjustable across a large frequency range of 0.5-7 THz and outstanding bandwidth tunability between 30 and 5000 GHz. Our approach is based on optical rectification of a temporally modulated chirped pump laser in an organic crystal. The scheme offers continuously tunability from single to a multi-cycle pulse exceeding 50 periods. The presented simple scheme is superior to other laser-based sources in view of efficiency, versatility and tunability as it offers a relative spectral bandwidth $\Delta\nu/\nu$ continuously selectable between 100% (multi-octave) and 2.5%. The scheme can be in principle extended towards linewidth of <1%. The results widen the potential of existing high-power lasers as sources of versatile THz radiation and opens new opportunities for nonlinear applications with multicycle pulses at MV/cm field strength for time-resolved investigations.

In the past several schemes for forming tunable narrowband THz radiation have been presented. The two main approaches are based on charged particle beams and lasers. Large-size electron accelerators with beam energy up to several GeV provide THz pulses with up to 100's of uJ by means of transition, edge or Terahertz free electron laser (T-FEL) radiation [12-16]. The two former sources offer only broadband THz radiation and the opportunity for tunable narrowband spectral emission is minor. By employing an electron bunch train a bandwidth of typically >20% could be demonstrated [14]. However, the polarization from this source is radial which is rather unfavorable for many experiments as it requires lossy polarization filtering to transform it to a well-defined linearly polarization state. Furthermore the user accessibility is restricted to such large scale sources as the primary usage of these accelerators is different. Dedicated T-FELs deliver radiation at tens of uJ pulse energy with a bandwidth variable between 0.2-5% in the range of ≈0.3-10 THz [17]. Undulator-based sources, such as the T-FELs, deliver multicycle fields with number of oscillation proportional to the undulator periods (typically 10-100) and are not suited for single-cycle THz radiation. As THz generation starts from noise the field from T-FELs does not offer a stable absolute phase. This is a particular burden for phase-sensitive experiments at sub-cycle resolution. Furthermore the synchronization with an external optical femtosecond laser required for pump probe experiments is strenuous and requires advanced technology.

Table-top sources delivering narrowband THz radiation have been realized by employing a molecular gas laser pumped with a powerful $CO_2$ laser or by optical down-conversion of short pulse laser using second-order $\chi^{(2)}$ optical rectification (OR) process in nonlinear crystals. While the former is limited in tunability to a set of specific molecular transitions the latter is typically used for broadband, single-cycle pulse production. However, routes towards a spectral confinement of THz radiation has been shown for OR-based sources by employing pump pulse shaping technique but the performance in view of spectral width,

tunability of the central frequency and brightness still deviates significantly from T-FEL based sources. Pulse shaping of the pump beam by spatial [18] or spectral [19] control results in narrow THz radiation tunable between 0.5-2 THz with a relative spectral bandwidth of typically 20-30%. Other schemes employ periodically-poled nonlinear crystals with a specially engineered multi-layer structure to generate THz at a fixed frequency [20]. Unfortunately the resulting THz pulse energy of typically <100 nJ is not sufficient to initiate nonlinear processes. Similar restrictions for THz pulse energy and field hold for narrowband THz sources based on beating two broadband chirped pump pulses either on a dipole antenna [21] via a fast oscillating current induced by the beat frequency or in nonlinear crystals [22-23]. Overall the performance for state of the art laser-based sources has been limited so far in view of central frequency tunability at narrow bandwidth, pulse energy and field strength. The scheme presented here overcomes all of these shortcomings extensively.

**Results**

The experiments were conducted on the Cr:Forsterite chirped-pulse amplification laser shown in Fig. 1 emitting 120 fs pulses at a wavelength of 1.23 μm and repetition rate of 10 Hz [24] in JIHT RAS. The laser provides up to 10 mJ pulse energy after compression. The laser spectrum of ≈22 nm (FWHM) is well suited for broadband and efficient optical rectification in the 400 um thick organic crystal DSTMS (4-N,N-dimethylamino

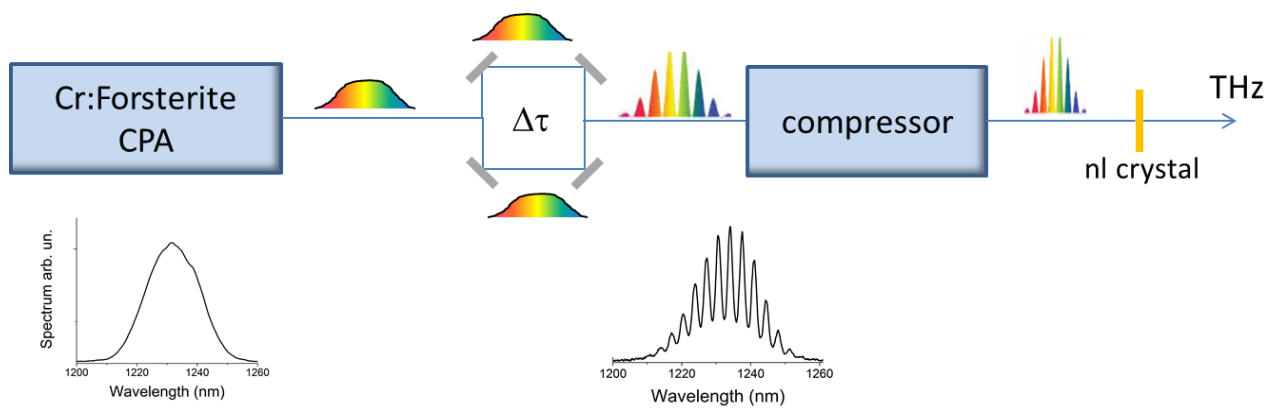

*Figure 1. Experimental layout. The amplified laser pulse from a Cr:forsterite CPA laser is split and delayed by means of a Mach Zehnder-type interferometer before re-compression to the few picosecond or femtosecond regime. The beat frequency of the two chirped pulses is selected by varying the delay Δτ. Shown as inset is the fundamental spectrum as well as an example of a frequency-beated spectrum, respectively, which forms a pulse train in the time domain. This pulse train is used for narrowband THz production in the nonlinear organic crystal.*

-4'-N'-methyl-stilbazolium 2,4,6-trimethylbenzenesulfonate) used in our studies [25]. The setup allows for the formation of a pulse replica by means of a Mach-Zehnder type interferometer introducing a second pulse at variable delay Δτ. As the two co-propagating replicas are stretched in time a frequency chirp beating occurs in case of temporal overlap. The considered linearly chirped optical pulse with a Gaussian envelope can be described as $E(t) =$ $E_0 exp(-\alpha t^2) exp\left(i(\omega_0 t + bt^2)\right)$ with $E_0$ the amplitude, $\alpha^{-0.5}$ the transform-limited pulse duration, $\omega_0$ the carrier frequency and b the chirp rate. The time intensity pattern of the beating pulses $I(t,\tau) = |E(t) + E(t+\tau)|^2 = I(t) + I(t+\tau) + 2\sqrt{I(t)I(t+\tau)}\cos(\omega_0\tau + b\tau^2 + 2b\tau t)$. The beating frequency $f=2\times b\tau$ gives rise to a modulated pump beam in the frequency domain which is easily modifiable by changing the delay τ and frequency chirp rate *b* of the pump laser. The highly modulated pulse train in the frequency domain exhibits a similar structure in time after stretching the pulse to several picoseconds, i.e. 10-100 fold longer than the transform-limited pulse duration. These modulated pulses are employed for multicycle THz production in DSTMS.

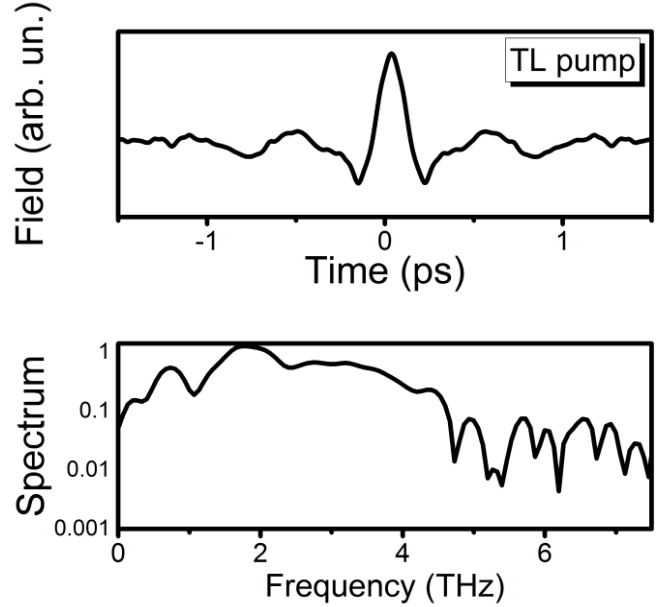

*Figure 2. Broadband THz radiation. Optical rectification of the transform-limited pump pulse in the organic crystal DSTMS and OH1 gives rise to a single-cycle THz field (shown in (a)) carrying a multi-octave spanning spectrum (b).*

The largest THz spectral output is achieved using the un-chirped pulse. In this configuration without frequency beating the temporal shape of the pump pulse is Gaussian. The resulting single-cycle THz pulse produced by OR in DSTMS shows an asymmetric field evolution and carries spectral content ranging from 0.1 to 7 THz (Fig. 2). Pumping with 10 mJ pulse energy at a fluence of 15 mJ/cm$^2$ results in a THz output of 280 μJ, corresponding to a 2.8% energy conversion efficiency and photon conversion efficiency of 330%. Albeit the Cr:Forsterite oscillator and amplifier system are not carrier envelope phase stabilized the THz pulses are phase-locked and carry an absolute phase which is constant for consecutive shots. This is expected from the optical rectification process where due to difference frequency generation between two intrapulse frequencies the phase offset is removed.

In order to gain control over spectral width and the central frequency the THz output the shape of the pump pulse is properly modified. This is achieved by chirping and delaying the two pump replicas, which introduces strong frequency beating. In first order the chirp defines the spectral bandwidth while the delay Δτ defines the central frequency. The broadband phase-matching range of the organic crystals allows for a continuous tunability of the center frequency between 0.5 and 7 THz, which is demonstrated in Fig. 3a. The resulting narrowband, multicycle THz pulses are generated with a chirp-and-delay setting specific to each central frequency in order to keep the relative bandwidth constant at ≈15% across the demonstrated tuning range. In this configuration the THz pulse energy reaches up to 20 uJ for frequencies around 2-3 THz where the TL spectrum shows the largest spectral density (Fig. 3b). The energy conversion efficiency of 0.2% is more than an order of magnitude higher than what has been reported in the past for laser-based narrowband THz sources. As expected from the bandwidth of the optical rectification in organic crystal the effective nonlinearity the THz pulse energy is frequency dependent and our observations confirm the frequency-scaling as a function of the spectrum provided by a transform-limited pump pulse (Fig. 3e, grey). We measured the energy carried by a narrowband THz pulse to scale in first order with the FWHM spectral bandwidth. The corresponding field strengths of the multi-cycle THz pulses are reported in Fig 3c. The largest field of 8 MV/cm is achieved around 4 THz center frequency and drops towards the edges due to the phase-matching issues (higher frequency range) and frequency-dependent increase of the diffraction-limited focus size (lower frequency range).

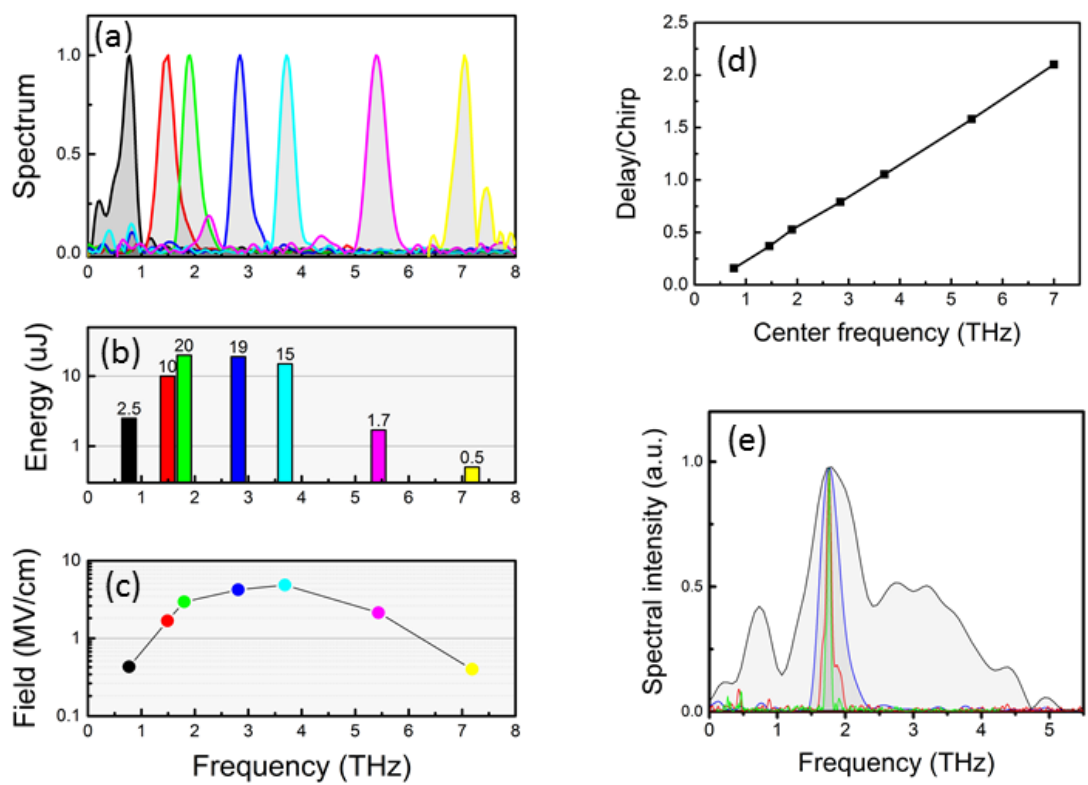

*Figure 3 Frequency and bandwidth tunability of strong-field THz pulses across the THz gap. (a) The scheme based on Cr:Forsterite pumped organic crystals provides an extreme coverage over almost 4 octaves, from 0.5 to 7 THz. Shown in (b) and (c) are the corresponding pulse energy and field strength of these narrowband pulses. The selected tunability at constant bandwidth is achieved by selecting the corresponding ratio between delay and chirp shown in (d). (e) The naturally broadband THz spectrum (grey) from the TRL pump pulse is dramatically narrowed by employing frequency beating of the pump pulse (blue, green, red). The bandwidth can be continuously reduced from multi-octave to <50 GHz (FWHM).*

Our scheme gives a straight-forward control on the spectral bandwidth by chirp and delay adjustment of the pump pulses. Shown in Fig 3e is the evolution of the bandwidth when the chirped pulse duration is increased from transform-limited 120 fs to 20 ps. The spectrum gets continuously reduced and results finally in a 45 GHz FWHM spectral line (green) at 1.8 THz, corresponding to a relative bandwidth of 2.5%. For this set of measurements the ratio between chirp and delay is kept constant in order to keep the resulting spectrum centered on the targeted 1.8 THz. We mention that, in general, for each central frequency chosen in the THz emitting range, spectrally narrowband radiation with a linewidth down to a few percent can be produced.

**Discussion**

Narrowband THz radiation in the low frequency THz part (0.1-1.5 THz) is of particular interest in quantum physics to selectively excite low-energy states with narrow linewidth. In order to provide spectrally-confined tunable source in this frequency range we optimized the generation conditions by employing a OH1 (2-(3-(4-Hydroxystyryl)-5,5-dimethylcyclohex-2-enylidene)malononitrile) organic crystal [26]. While THz emission from DSTMS is affected by a strongly absorbing phonon resonance at 1 THz the OH1 crystal emits broadband radiation centered at 1 THz when pumped with a transform-limited pulse from Cr:Fosterite laser. Shown in Figure 4 is the continuous tuning range of the OH1 source over almost three octaves when pumped with our modulated pump configuration. The central frequencies between 0.28 and 1.7 THz are achieved by modifying chirp/delay ratio accordingly. The minimum measured bandwidth of 30 GHz is sufficiently narrow to drive for the first time selectively a quantum state with a table-top THz source. The conversion efficiency across the frequency conversion range provided by OH1 is ≈0.02% and reaches up to 5 uJ (central frequency at 1.1 THz) in our experiment. The corresponding field strength is of hundreds MV/m at diffraction limited focal spot. The demonstrated field strength is almost two orders of magnitude beyond the state of the art of laser-based sources and comparable what is achieved at large-size T-FEL facilities. In principle even narrower bandwidth could be achieved in our scheme by employing pump beam with a super-Gaussian like temporal shape by means of infrared spectral shaper. However, such a shaper was not available for this proof of principle. Finally our scheme allows for a variety of intrinsically jitter-free multicolor pump and optical probe schemes while offering THz pulse shaping capabilities covering single-cycle, GV/m pulses as well as many-cycle pulses with hundreds MV/m field strength. Scaling towards larger field strength and pulse energy and subsequently larger field strength is straightforward by employing higher pump energy and larger organic crystals.

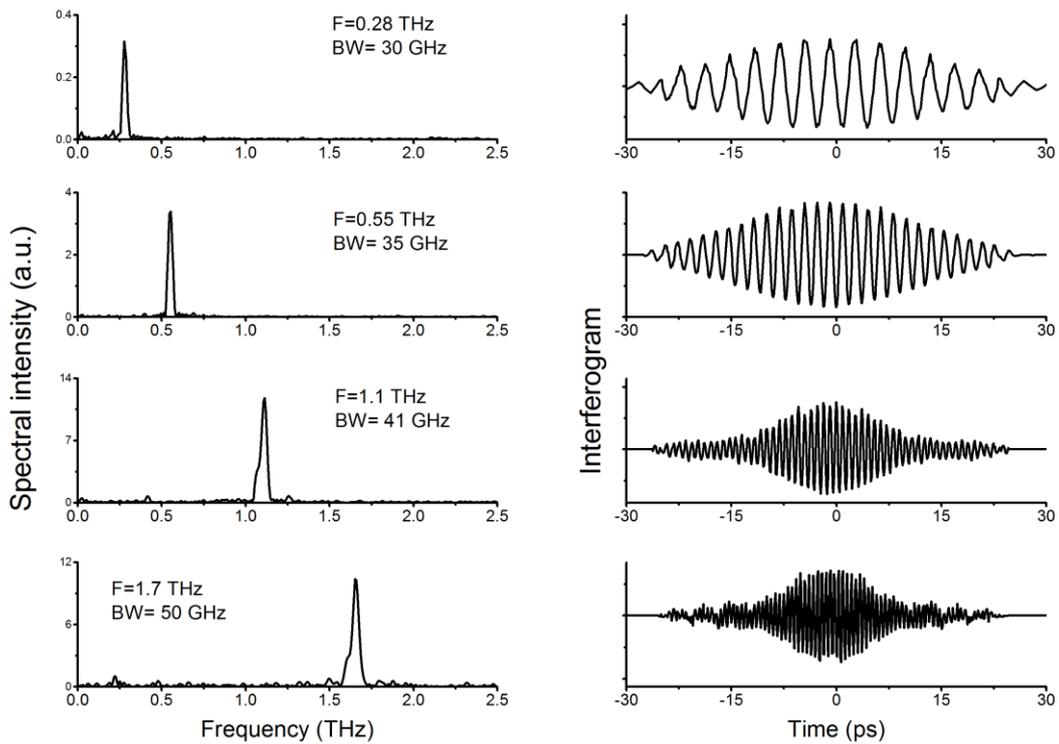

Figure 4 Narrowband THz radiation continuously tunable across the 3-octave spanning low THz frequency range (0.2-1.6 THz) by employing chirp-and-delay pulses for OR in the organic crystal OH1. The corresponding pump beating pattern is shown as well (right column).

In conclusion we demonstrated a versatile high-peak power THz source capable to deliver both multi-octave spanning as well as narrowband THz radiation continuously tunable in central frequency and bandwidth across 5 octaves. The pump pulse intensity profile modulated by frequency beating of two chirped pulse replica is optically rectified in highly efficient organic crystal DSTMS and OH1. The THz output linewidth is tunable across the frequency range of 0.2-7 THz and offers a record small bandwidth of 30 GHz. The highly efficient organic crystal-based THz generation provides up to 300 μJ at broadband and 5 uJ in a 30 GHz narrowband mode at a conversion efficiency of 2.8% and 0.02%, respectively. The unprecedented field strength of GV/m (single-cycle) and hundreds of MV/m (multi-cycle) opens new scientific pathways for selective and nonlinear excitation of low energy modes in condensed matter. Moreover the presented source disclose new prospective for the realization of compact electron accelerators and for advanced imaging applications.

## Methods

For our THz source, we use an high-energy Cr:forsterite laser with pulse duration of 120 fs to pump large-size organic crystals OH1 and DSTMS (thickness of 440 μm and diameter of 6 mm) at 1.23 μm wavelength. The spectrum of the generated THz pulse is centered around 4 THz. Terahertz pulse detection was performed using a Michelson-based autocorrelator
.

## Author contributions:

CPH and CV conceived the experiment. CV, AO, OVC and CPH performed the experiment. CPH wrote the manuscript, with support from CV. All authors contributed to the discussion and analysis of the results.

## Acknowledgements

This work was supported by Swiss National Science Foundation project no.: IZLRZ2_164051 and by the Ministry of Education and Science of the Russian Federation, project no. 14.613.21.0056, RFMEFI61316X0056.